\documentclass[aip,reprint]{revtex4-1}

\usepackage{amsmath,amsfonts,amsthm,bm} 

\usepackage{amssymb}
\usepackage{graphicx}
\usepackage{soul} 
\usepackage{xcolor} 

\usepackage{url} 
\usepackage{hyperref}

\usepackage{float}

\usepackage{tabularx}

\usepackage{mdframed}
\usepackage{tcolorbox}

\usepackage{caption}  
\usepackage{lipsum}   
\usepackage{ragged2e}

\newtcolorbox{myshadedbox}[2][]{colback=blue!10!white, colframe=blue!50!black, fonttitle=\bfseries, title=#2, boxrule=0.5mm}

\newcounter{boxnumber}
\newcommand{\boxtitle}[1]{%
   \refstepcounter{boxnumber}
    \textbf{Box \theboxnumber: #1}
}

\draft 

\begin{document}

\title{Programmable metasurfaces for future photonic artificial intelligence}

\author{Loubnan Abou-Hamdan}
\thanks{These authors contributed equally to this work.}
\affiliation{Department of Physics, Colorado School of Mines, 1523 Illinois Street, Golden, CO 80401, USA}

\author{Emil Marinov}
\thanks{These authors contributed equally to this work.}
\affiliation{Department of Physics, Colorado School of Mines, 1523 Illinois Street, Golden, CO 80401, USA}
\affiliation{Université Côte d’Azur, CNRS, CRHEA, Rue Bernard Gregory, Sophia Antipolis, 06560, Valbonne, France}
\author{Peter Wiecha}
\affiliation{LAAS, Université de Toulouse, CNRS, Toulouse, France}

\author{Philipp del Hougne}
\affiliation{Université de Rennes, CNRS, IETR - UMR 6164, F-35000, Rennes, France}
\author{Tianyu Wang}
\affiliation{Department of Electrical and Computer Engineering, Boston University, Boston, MA 02215, USA}

\author{Patrice Genevet}
\thanks{Corresponding author: patrice.genevet@mines.edu}
\affiliation{Department of Physics, Colorado School of Mines, 1523 Illinois Street, Golden, CO 80401, USA}

\begin{abstract}
\textbf{Photonic neural networks (PNNs), which share the inherent benefits of photonic systems, such as high parallelism and low power consumption, could challenge traditional digital neural networks in terms of energy efficiency, latency, and throughput. However, producing scalable photonic artificial intelligence (AI) solutions remains challenging. To make photonic AI models viable, the scalability problem needs to be solved. Large optical AI models implemented on PNNs are only commercially feasible if the advantages of optical computation outweigh the cost of their input–output overhead. In this Perspective, we discuss how field-programmable metasurface technology may become a key hardware ingredient in achieving scalable photonic AI accelerators and how it can compete with current digital electronic technologies. Programmability or reconfigurability is a pivotal component for PNN hardware, enabling in situ training and accommodating non-stationary use cases that require fine-tuning or transfer learning. Co-integration with electronics, 3D stacking, and large-scale manufacturing of metasurfaces would significantly improve PNN scalability and functionalities. Programmable metasurfaces could address some of the current challenges that PNNs face and enable next-generation photonic AI technology.}
\end{abstract}

\maketitle 


\section*{Introduction}

\noindent
Since the 1960s, computational demand has doubled approximately every two years, and fortunately, it has been met with an apposite doubling of computational power, following Moore's empirical law. However, in recent years, the computational and energy demands of digital neural networks have rapidly increased, doubling approximately every three months\cite{openai,makarenko2023photonic}. The current boom in artificial intelligence (AI) technology is pushing technology to a critical point at which the scaling laws governing digital computing hardware are reaching their physical limits and approaching saturation. This has resulted in unprecedented energy consumption imposed by the high computational demands of AI technology (see Box~\ref{box 1}), which may make it unsustainable in the near future.

One promising solution to reduce the energy consumption of neuromorphic computing is using photonic neural networks (PNNs). They are an emerging hardware alternative to digital neural networks in which computationally expensive matrix-vector multiplications are implemented optically. Calculations in PNNs are performed using arrays of pixels or meta-atoms with tailored electromagnetic properties. PNNs may also include an optical or electronic nonlinear activation function to enhance network depth. This kind of architecture exploits the inherent diffraction property of electromagnetic waves propagating across optical interfaces so that calculations are performed directly on the lowest hardware abstraction level. In contrast, digital neural networks rely on high-speed routing of binarized electric signals between trillions of transistors on which various layers of computational abstraction are built. 

The use of PNNs is especially advantageous in terms of energy efficiency, latency and throughput\cite {mcmahon2023physics, huang2024photonic}. This is a direct consequence of the large bandwidth of light, its nearly dissipationless propagation in lossless media (that is, without heat generation, which is the main limiting factor of current digital hardware), and its high spatial and frequency parallelism. The propagation of light through optical components (lenses and thin or volumetric scatterers) naturally performs linear complex-valued mathematical operations, such as forward or inverse Fourier transforms\cite{goodman2005introduction} and, more generally, matrix multiplication\cite{cheng2021photonic, zhou2022photonic, nikkhah2024inverse}.
These mathematical operations account for the majority of the computational costs and memory requirements of digital neural networks\cite{goodfellow2016deep}. As light can naturally execute such linear complex-valued mathematical operations, it has been used to develop optical computing structures\cite{silva2014performing} capable of performing compact wave-based analogue \cite{goh2022nonlocal} and parallel computing \cite{camacho2021single,sol2022meta}, and solving equations \cite{mohammadi2019inverse, cordaro2023solving}. 

In recent years, significant strides have been made in the field of photonic artificial intelligence\cite{ferreira2017progress, wetzstein2020inference}, and PNNs have been implemented using photonic-integrated circuits (PICs)\cite{shen2017deep, feldmann2019all,feldmann2021parallel, wang2022integrated, pai2023experimentally}, optical cavities\cite{momeni2023backpropagation,xia2023deep}, multi-mode fibres\cite{rahmani2020actor,momeni2023backpropagation} and spatial light modulators (SLMs)\cite{zhou2021large, wang2023image}. Optical matrix-vector multipliers (MVMs) have been successfully implemented via linear multi-layer diffractive elements\cite{lin2018all, qian2020performing, luo2022metasurface, luo2022cascadable, ding2024metasurface} (see Box~\ref{box 2}), in architectures that lack reconfigurability. However, to develop PNNs in which non-stationary use cases, fine-tuning, and transfer learning are implemented, programmable light modulation is required. As a result, the challenge of conceiving massively scalable programmable PNNs, in which the same hardware can be used as a computing component or an AI accelerator, remains an ongoing issue that has yet to be addressed. The scalability of PNNs is vital in this regard, as the benefits in energy efficiency and throughput gained via optical computation become appreciable for large-scale PNNs\cite{mcmahon2023physics}. As a result, the development of scalable photonic AI hardware has recently become an active area of research\cite{hamerly2019large, huang2020demonstration, onodera2024scaling}.

Most attempts at producing programmable PNNs have relied on photonic integrated circuits. Despite their maturity, they intrinsically present scalability issues due to the large footprint of individual nodes in two-dimensional (2D) architectures. For example, a state-of-the-art integrated photonic neural network with 132 learnable parameters has recently been implemented on an on-chip 3-layer architecture\cite{bandyopadhyay2024single}. However, scaling such a system to a 1 billion parameter neural network would result in a chip surface of around $3.4$~m$^{2}$, which is unrealistic. (This figure was calculated from the number of Mach-Zehnder modulators (45) necessary to implement the network's 132 learnable parameters. The size of each Mach-Zehnder modulator was approximated as $50\times 200$~$\mu\text{m}^{2}$. In general, the exact spatial footprint of a node in a photonic integrated circuit depends on various factors, including operation wavelength, material, and integration level.) A tentative solution to resolve this 2D PIC architecture limitation has been proposed using three-dimensional (3D) manufacturing to print 3D-integrated PNNs\cite{moughames2020three, dinc2020optical}. This approach is potentially interesting but still lacks reconfigurability and fabrication simplicity. Another approach involves continuously tuned photo-refractive materials\cite{onodera2024scaling} that could potentially exhibit a lower 2D footprint. Nevertheless, the effective pixel size is currently limited to around $50\, \mu\text{m}^{2}$ because of carrier diffusion. Wavelength multiplexing with Kerr micro-comb sources has also been used to improve the scalability of 2D planar architectures\cite{xu202111}.

The scalability limitations of 2D planar devices can be circumvented by adopting a 3D free-space approach.  However, most of the devices involved in free-space setups are expensive and bulky, suffer from inherently slow response times, and mostly work only in reflection. As such, there is a need for devices that benefit from the free-space approach without the shortcomings of conventional free-space modulators. The ideal device for such an application would be a programmable, cost-effective optical modulator that can provide significantly higher density, efficiency, and modulation speed. 
Reconfigurable optical metasurface technology seems a natural solution \cite{yu2011light, aieta2015multiwavelength, genevet2017recent} for an alternative scalable PNN technology (note that the terms "reconfigurable" and "programmable" are used interchangeably herein). Consider a metasurface with a pixel size of $300 \times 300$~nm$^{2}$: this would allow the implementation of 1 billion parameters for a PNN within a surface as small as 0.9~cm$^{2}$. The overall dimensions of a metasurface optical chip containing driving electronics and additional packaging will nevertheless increase the chip footprint, but similar considerations also apply to other optical computing technology. The issues of electrical contacting, driving, and fabrication difficulty are tackled in more detail in the Open Challenges section. We also refer the reader to Box~\ref{box 3} for a comparison of performance metrics of electronic and photonic computers. 

\begin{figure*}[t!]
\centering
\includegraphics[width=\linewidth]{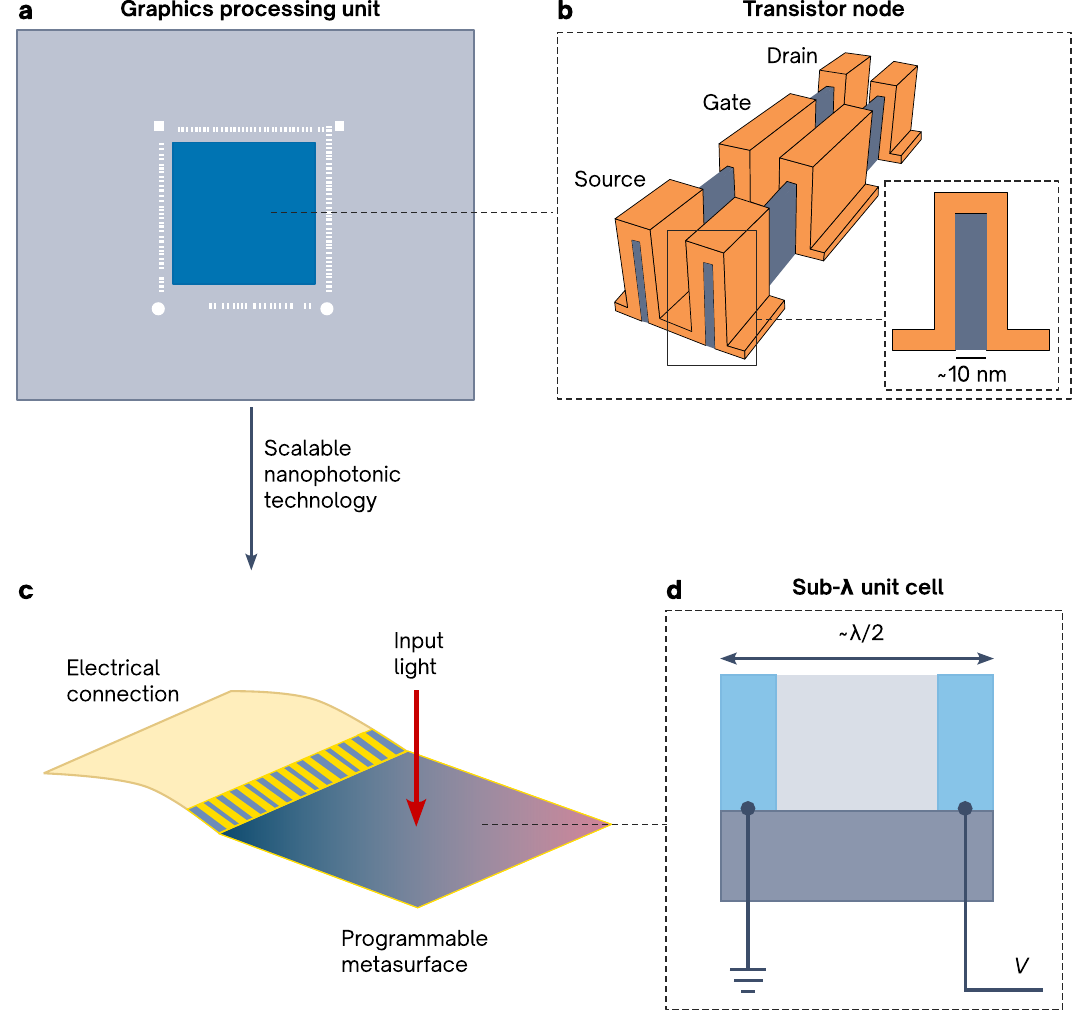}
\caption{\textbf{An illustration of the perspective programmable metasurface technology for optical computation.} The electrically connected programmable metasurface (panel \textbf{c}) consists of subwavelength voltage-actuated pixels (panel \textbf{d}). Due to the metasurface’s reconfigurability, structural nonlinearities can be implemented. Such technology could mimic electronic processors, such as a graphics processing unit (GPU, panel \textbf{a}), which consists of billions of transistor nodes (panel \textbf{b}).}
\label{fig1}
\end{figure*}

In this Perspective, we briefly review current state-of-the-art photonic neural networks and discuss the programmable optical metasurface solutions that are currently available and that can be adapted for next-generation programmable metasurface neural network technology (see Fig.~\ref{fig1}). The combination of subwavelength resolution and optical neuron reconfigurability would give programmable metasurface neural networks the versatility that is necessary to update the system with new data at any time, thus addressing various inference tasks using the same hardware. Such plasticity would propel photonic AI architectures to a level in which they play both the role of hardware and software\cite{laydevant2024hardware}, with training and inference realized simultaneously. Optical plasticity would enable high-speed continual learning with non-stationary datasets. Note that continual learning refers to the capacity of a neural network to learn and implement new capabilities while maintaining previously learned functionalities. The programmable nature of the neural network also allows nonlinear activation to be implemented in the form of structural nonlinearities.

\section*{The role of nonlinearities}

\noindent
To achieve deep learning capabilities, that is, the ability to learn and model complex and nonlinear relationships in data, neural networks generally need nonlinear activation. Nonlinearities enable a neural network to build a feature detection hierarchy in which successive layers build increasingly abstract representations of the input data. This has proven essential for an accurate approximation of complex functions. Its mathematical proof uses the universal approximation theorem\cite{cybenko1989approximation, hornik1989multilayer, huang2006universal}, which is a foundational result in the theory of artificial neural networks. The theorem asserts that a feedforward neural network with at least one hidden layer and nonlinear activation (typically an arbitrary bounded and non-constant activation function\cite{hornik1991approximation}) can approximate any continuous function on a closed and bounded subset of $\mathbb{R}^n$ to an arbitrary degree of accuracy, provided there are enough neurons in the hidden layer.

Although nonlinearities are routinely used in optics for frequency generation processes and to realize optical switches, PNNs involving nonlinear activation functions are rare, typically owing to weak conversion efficiencies. In contrast to a deep digital neural network (Box~\ref{box 2}, panel \textbf{a}), the diffractive multilayer structure shown in Box~\ref{box 2}, panel \textbf{b}, consists of a purely passive and linear architecture that can be mathematically collapsed to a single matrix-vector multiplication. If we consider such a diffractive system consisting of $L+2$ layers (where the $0^{\text{th}}$ layer is the input layer and layer $L+1$ is the optical read-out layer) each composed of $N^{2}$ complex-valued pixels (that is the 0th layer encodes information in $N\times N$ pixels), the network's input-output relationship is described by the matrix-vector multiplication $\mathbf{x}_{L+1} = M \mathbf{x}_{0}$, where $\mathbf{x}_{l}$ ($l\in[0,L+1]$) is an $N^{2}$-dimensional vector and $M$ is an $N^{2}\times N^{2}$ matrix\cite{wei2018comment}. Note that each layer of the actual physical hardware consists of $N\times N$ pixels; however, since the input information was flattened into an $N^{2} \times 1$ vector, this results in a matrix–vector multiplication in which the full optical system is represented by an $N^{2} \times N^{2}$ matrix.  Furthermore, the matrix $M$ is independent of the number of layers $L$, so that such a system can be regarded as a complex linear mode-matching operator trained to map linear relationships between the input and output. Nevertheless, the accuracy of such linear wave systems performing linear tasks may be improved by increasing their number of layers (increasing the number of neurons or tuning parameters used to minimize errors between inputs and outputs)\cite{lin2018all, mengu2018response, raeker2019compound, nahmias2019photonic, kulce2021all}, suggesting that the number of layers acts as an adjustable parameter that may improve the network's accuracy.\\

 It is now well established that certain simple machine learning tasks (such as the MNIST dataset\cite{lecun2010mnist} classification task) can be solved with linear diffractive devices whose performance can be enhanced with the addition of layers, as explained above. Multilayer linear diffractive models, however, fail to effectively capture nonlinear relationships between input and output data\cite{marcucci2020theory, brunner2025roadmap} (see Box~\ref{new box}). As a result, the investigation of nonlinearities in the context of photonic neural networks is an exciting research direction that can lead to improved performance and enhance the breadth of neural network technology, particularly when implementing complex machine learning tasks. Note that the study of nonlinear activation in photonic neural networks is still a topic of intense discussion (see ref. \cite{brunner2025roadmap}). \\

When it comes to PNNs, implementing nonlinearities has been challenging. Optical nonlinear effects are generally extremely weak, which makes it necessary to work with high field intensities. Sophisticated engineering, such as phase matching, is often needed to harness these effects (see supplementary information, section~I). Interestingly, however, nonlinear mappings can be opto-electronically implemented using purely linear wave systems by exploiting their so-called structural nonlinearity. In a neural network, structural nonlinearities can be induced by reconfiguring some layers based on the input signal information (see below). This approach, which promises to realize scalable PNNs with embedded nonlinearities, only applies to reconfigurable optical layers, such as programmable/reconfigurable metasurfaces. 
\vspace{0.25cm}

\subsection*{Structural nonlinearity}

\noindent
Structural nonlinearities in optics refer to nonlinear dependencies of the transfer function of a linear optical system on its structural configuration. They are compatible with linear wave propagation and do not require nonlinear light–matter interactions. To exploit structural nonlinearities, the input data must be encoded into the optical system's structural configuration as opposed to encoding them into the impinging wavefront. Meanwhile, the output data are extracted from measurements of the optical system's transfer function. Reconfigurable metasurfaces are ideally suited to exploit structural nonlinearities since their configuration (amplitude and phase values of each pixel) can be chosen as a function of the input data. The nonlinear transformation from input data to output data is then realized by an optoelectronic system comprising the linear optical system and the reconfigurable metasurface's control circuitry.

In more technical terms, a linear wave system linearly maps an input wavefront $\mathbf{x}$ to an output wavefront $\mathbf{y} = \mathbf{H}\mathbf{x}$. If the linear wave system is parametrized, for example by a programmable metasurface with configuration $\mathbf{c}$, then the input-output mapping $\mathbf{H}$ from $\mathbf{x}$ to $\mathbf{y}$ is still linear, but $\mathbf{H}$ depends nonlinearly on $\mathbf{c}$. More specifically, this nonlinear dependence can be traced back to a matrix inversion in physical models of programmable linear wave systems; this matrix inversion can also be cast as multiple nested infinite sums of matrix powers, physically representing paths that involve multiple reflections owing to recurrent scattering~\cite{rabault2024tacit}. One can therefore implement a nonlinear mapping from the input information to the output information through a perfectly linear wave-interfering system by encoding the input information into $\mathbf{c}$ rather than into $\mathbf{x}$, and extracting the output information from $\mathbf{H}$ (by measuring $\mathbf{y}$ for several distinct and known $\mathbf{x}$) instead of $\mathbf{y}$. In systems that are well approximated as mainly forward-scattering, in which mutual coupling and multiple scattering effects are negligible, one can artificially construct a truncated version of the aforementioned infinite sum with multiple programmable layers\cite{yildirim2023nonlinear}. 

The advantage of reconfigurability and the possibility of inputting arbitrary information into $\mathbf{c}$, including replicas or arbitrary transforms of the input wavefront, may potentially produce complicated non-linear functions. This is an optoelectronic approach as the input information is typically encoded optoelectronically in the metasurface configuration rather than in an optical signal. Note that structural nonlinearity can be induced in a purely optical setting, such as in an optical cavity\cite{xia2023deep}, but encoding different inputs requires some reconfigurability, which is typically achieved optoelectronically or with optical bias. To ensure that the speed of optical computing is unhindered by the optoelectronic reconfiguration process, metasurfaces with high-speed reconfigurability (with switching speed exceeding or on the order of the information feeding technique) become critical. Currently, the highest switching speed in metasurfaces is on the order of a few gigahertz, achieved using the electro-optic effect (see Table~\ref{table1}), which is sufficiently fast to implement structural nonlinearities in fast high-performance systems.  \\

Although the idea of encoding inputs into the configuration of a programmable metasurface that parameterizes a linear wave system has been around for a few years~\cite{del2018leveraging}, it has only recently been used to provide nonlinear mappings in PNNs~\cite{momeni2023backpropagation}. This concept was initially explored in the microwave regime~\cite{del2018leveraging,momeni2023backpropagation} and has now also been investigated at optical frequencies~\cite{wanjura2023fully, yildirim2023nonlinear,xia2023deep, li2024nonlinear}.

\section*{Photonic neural network training}

\noindent
Before a neural network can perform inference, that is, predict an output from a given input, it must first undergo a training phase in which its parameters are optimized for a specific task. This task is implicitly defined by the training set --- a large dataset that the initially untrained network evaluates. During this training process, the model's prediction error is quantified, for example, with a mean-square error loss function, and minimized by adjusting the model's weights. Several training schemes to train photonic neural networks\cite{momeni2024training} exist. We illustrate the most prevalent ones in Fig.~\ref{fig2} and briefly outline them below (see supplementary information, section~II for a more detailed discussion).

\subsection*{In silico training}
The most basic training scheme is in silico training (Fig.~\ref{fig2}a), in which a digital twin model, which replicates the physical diffractive neural network, is implemented on a computer and trained for a specific task. It typically uses error back-propagation learning\cite{rumelhart1986learning}, in which errors are back-propagated in the network and its parameters are optimized using stochastic gradient descent \cite{kingma2014adam}. This training scheme has been successfully implemented to design and fabricate linear diffractive MVMs that can accurately perform imaging and classification tasks in the terahertz regime (see Fig.~\ref{fig2}a). However, the most apparent drawback of this training method stems from the unavoidable discrepancies between the simulated model and the experimental system, which arise from fabrication errors and other experimental uncertainties that directly impact performance\cite{goi2022impact}. However small, these errors ultimately give rise to a simulation-reality gap that increases as a function of the number of layers and training steps\cite{wright2022deep}.

\subsection*{Physics-aware training}

\noindent
Comparing realistic experimental measurements with approximated numerical models can cause problems. To mitigate this issue, a physics-aware training scheme has been proposed. The difficulty of building a physical computing system (such as a PNN) is creating a computer model that perfectly mimics realistic systems. Real-world hardware invariably has tiny flaws or patterns of noise that may change over time. Physics-aware training (PAT) refers to a set of techniques designed to bridge the simulation-reality gap. PAT renders the training process more "aware" of the actual physical system by using real-time data directly from a feedforward pass through the physical system rather than relying on predictions from a computer model. This ensures that adjustments to parameters during training accurately reflect the system's actual physical behavior.

\begin{figure*}[ht!]
\centering
\includegraphics[width=0.6\textheight]{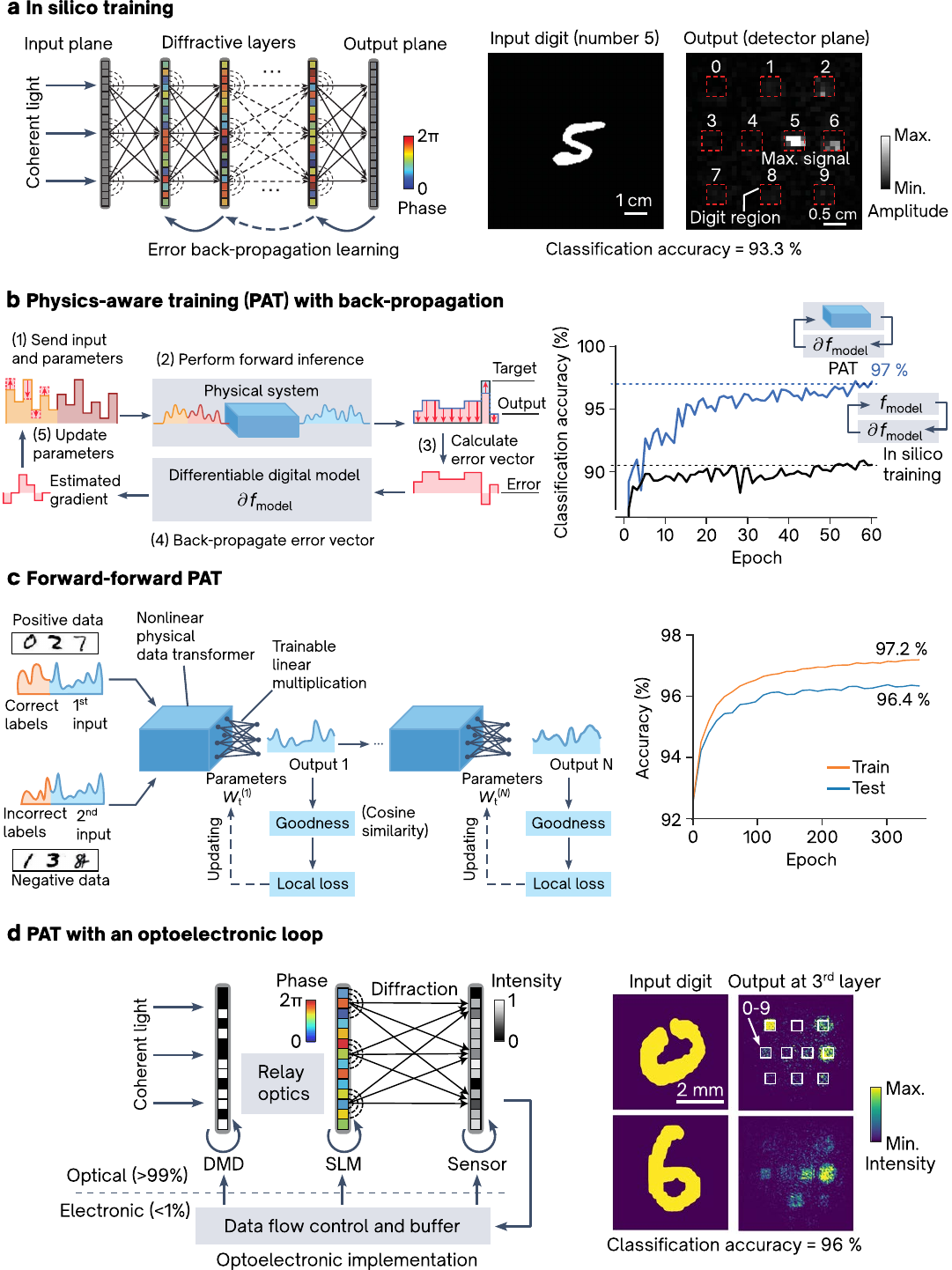}
\caption{\textbf{Illustration of training schemes used in free-space photonic neural networks.} \textbf{a}, In silico training performed on a digital twin of the physical diffractive neural network that is trained via error back-propagation. \textbf{b}, Physics-aware training (PAT) in which the forward pass is performed on the physical neural network, whose output is then used to compute the loss function and the back-propagated gradients in a digital model (denoted by $f_{\mathrm{model}}$), which are then used to update the physical system with the optimized parameters. \textbf{c}, Forward-forward PAT in which a local 'goodness' loss function is computed and minimized at the output of each layer following two forward passes, one in which positive data are input with correct labels and another in which negative data are input with incorrect labels. \textbf{d}, PAT with an optoelectronic loop in which a spatial light modulator (SLM) modulates the input from a digital micro-mirror device (DMD) in phase. A sensor reads out the output of the SLM, which is then fed back as an input for the DMD. Reiterating this feedback process several times mimics a multilayer diffractive neural network. The physical output at each layer is used to optimize the phase values of the SLM via back-propagation on a digital model. It should be noted here that this method can be used with any phase-modulating element other than an SLM. All experimental results shown are for a digit classification inference task. Panel a adapted from \cite{lin2018all}, panel b adapted from\cite{wright2022deep}, panel c adapted from \cite{momeni2023backpropagation}, panel d adapted from \cite{zhou2021large}. }  
\label{fig2}
\end{figure*}

PAT uses either error back-propagation\cite{wright2022deep} (Fig.~\ref{fig2}b) or forward-forward training\cite{hinton2022forward} (Fig.~\ref{fig2}c). In error back-propagation physics-aware training, the forward pass is implemented in the physical layers of the neural network, whereas error back-propagation and parameter optimization are performed digitally. Forward-forward physics-aware training\cite{oguz2023forward, momeni2023backpropagation} avoids back-propagation altogether by solely relying on forward passes through the physical neural network. In this approach, each input wavefront contains an input signal paired with a predefined "correct" or "incorrect" label. These two types of input wavefronts are referred to as "positive data" and "negative data", respectively (see Fig.~\ref{fig2}c). Each PNN layer is then optimized to maximally distinguish between positive and negative data. The loss function is thus designed so that a layer's output scores high "goodness" when processing positive data and low "goodness" when processing negative data\cite{momeni2023backpropagation}. To minimize this loss function, one has to adjust the layer’s parameters such that it produces outputs that better distinguish correct from incorrect labels.

Both types of physics-aware training have been used to produce photonic neural networks that can perform classification tasks with higher accuracy than in silico-trained neural networks (see Fig.~\ref{fig2}b and c). Spatially symmetric neural networks have also recently been shown to support fully forward error propagation\cite{xue2024fully}, which is an alternative approach to classical back-propagation that can be implemented entirely in the optical domain. Additionally, physics-aware back-propagation training has been implemented in a neural network with an optoelectronic feedback loop (see Fig.~\ref{fig2}d). This neural network is constructed from only one physical layer that mimics a multilayer network through repeated optoelectronic re-injection. Interestingly, this re-injection strategy leads to embedded nonlinearities as the re-injected signal is the intensity, which is read out by a sensor at the output layer\cite{zhou2021large}. In this case, the nonlinearity arises because the detected signal at the sensor (the field intensity) is a quadratic function of the electric field.

Physics-aware training represents a vital learning tool by which photonic neural networks can achieve optimal performance. However, such a training scheme requires real-time reconfigurability of the network parameters --- a task for which programmable metasurface neural networks are ideally suited. The optimal type of physics-aware training can vary from one system to another, which warrants a re-evaluation of learning methods in the context of PNNs. It is evident that the trend in photonic networks so far has relied heavily on ad hoc implementations of existing learning algorithms designed for digital computers (for example, error back-propagation). Nevertheless, these algorithms may be ill-suited to PNNs and may fall short of taking full advantage of photonic systems. Further inquiry into physics-based learning schemes in the photonic domain is, therefore, necessary and could lead to intriguing possibilities, such as physical self-learning machines\cite{lopez2023self}.

\section*{Programmable optical metasurfaces}

\begin{figure*}[ht!]
\centering
\includegraphics[width=\linewidth]{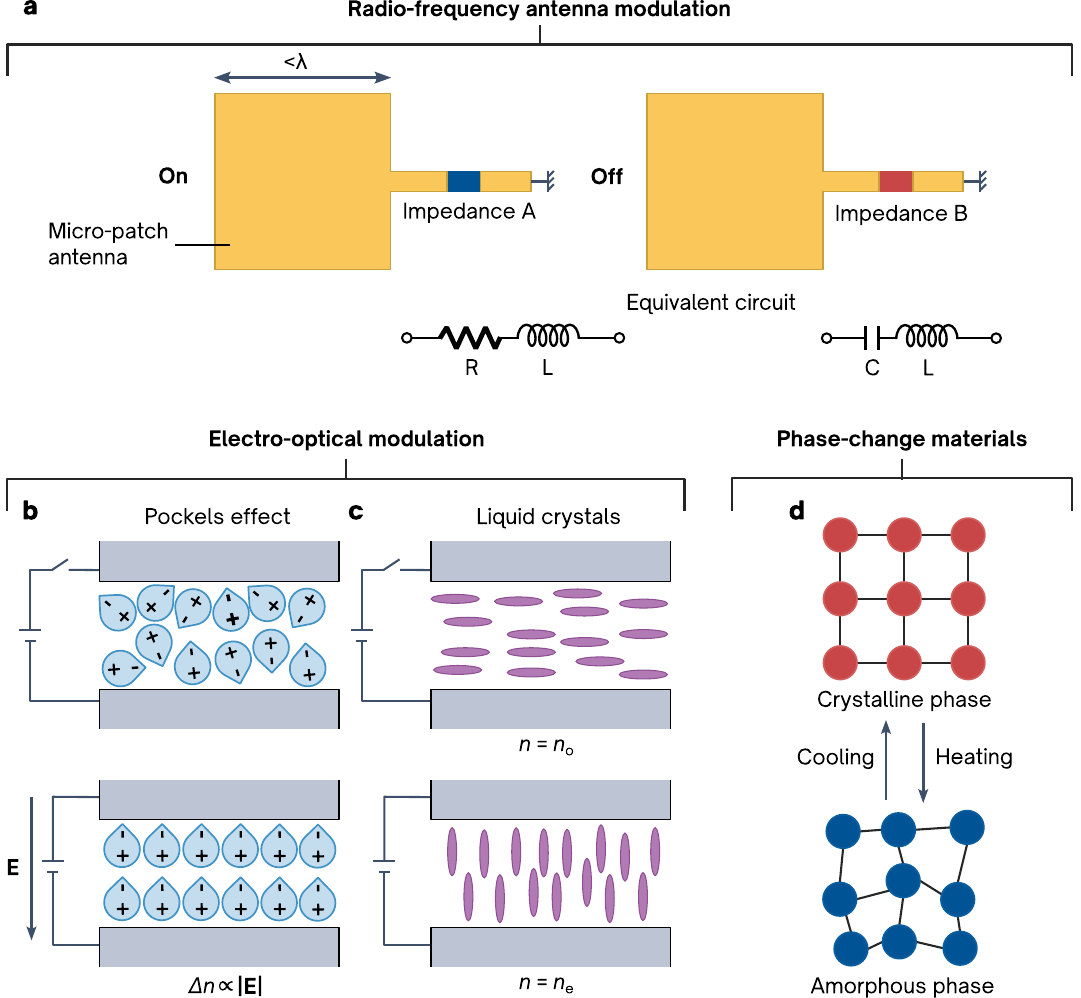}
\caption{\textbf{Physical mechanisms used to reconfigure electromagnetic fields in the radio-frequency and optical domains.} \textbf{a}, Radio-frequency reconfigurable metasurfaces rely on the modulation of the reflectivity of an array of sub-wavelength patch antennas made from a conducting material. Each antenna is connected to a lumped impedance (either a PIN diode or a varactor diode) that can be electrically modulated between a low-resistance state (impedance A, modelled primarily by an RL circuit) and a high-resistance or capacitive element (impedance B, modelled by a CL circuit). This modifies the antenna's impedance and modulates the reflection of electromagnetic waves from the antenna.~\cite{sievenpiper2003two, kamoda201160} \textbf{b}, The electro-optic Pockels effect\cite{karvounis2020electro, benea2021electro, weigand2021enhanced, zheng2024dynamic} can induce an electrically modulated refractive index in a material consisting of non-centrosymmetric molecules (represented by teardrop-shaped dipoles) to tune light waves. \textbf{c}, Liquid-crystal molecules in a unit cell consisting of two parallel electrodes across which an electric field is applied reorient themselves such that they are aligned along the field lines. This results in a modulation of the refractive index of the liquid-crystal from the ordinary ($n_{\text{o}}$) to the extraordinary ($n_{\text{e}}$) refractive index, effectively modulating electromagnetic waves traversing the liquid-crystal unit cell.~\cite{komar2017electrically, komar2018dynamic, kowerdziej2019ultrafast, sun2019efficient, li2019phase, dolan2021broadband} \textbf{d}, Phase-change materials can be switched from an amorphous to a crystalline state (and vice versa) through the application of a stimulus, such as the modulation of the material's temperature, thus modulating the material's refractive index\cite{huang2016gate, nicholls2017ultrafast, lewi2017ultrawide, wang2021electrical, zhang2021electrically, cueff2021reconfigurable, king2024electrically}. }  
\label{fig3}
\end{figure*}

\noindent
Programmable metasurfaces with exceptional beam-steering capabilities were first developed in the radio-frequency regime, using different terminology such as "tunable impedance surfaces", "reconfigurable reflect-arrays"~\cite{sievenpiper2003two, kamoda201160} (see Fig.~\ref{fig3}), "digital metasurfaces" or "spatial microwave modulators"~\cite{cui2014coding, kaina2014shaping}. These concepts have also been transposed to the terahertz regime~\cite{watts2014terahertz, wu2020liquid, chen2022electrically, lan2023real}.  

 More recently, reconfigurable optical metasurfaces\cite{mikheeva2022space, gu2023reconfigurable} have become an active area of research. However, device operation in the optical regime requires fundamentally different light modulation mechanisms than those operating at microwave or terahertz frequencies. Several modulation schemes based on various physical mechanisms have been proposed (see Fig.~\ref{fig3}b-d), but they all rely either on the modulation of the active medium's refractive index or mechanical movement.

We underscore here that for PNN systems to meet the standards set by digital computers and to address the challenges faced by such technology, they have to achieve several key performance metrics, most importantly bias type, modulation bandwidth, cyclability, and power consumption. In particular, optical hardware needs to achieve high modulation bandwidth (up to the gigahertz range), high optical efficiency (ideally close to 100\%), low power consumption (preferably biased electrically with bias voltages $\in [0,10]$~V), and high cyclability ($> 10^{10}$ cycles). To facilitate their integration with existing digital hardware, PNNs must also be stable from room temperature up to approximately $100^{\circ}$C. Finally, the reconfigurable optical hardware must be compatible with mass-manufacturing technologies. In this section, we review the light modulation schemes that satisfy some of these key performance metrics (see Table~\ref{table1}).

\subsection*{The electro-optic effect}

\noindent
So far, the fastest switching speed in reconfigurable optical metasurfaces has been demonstrated using the electro-optic Pockels effect (Fig.~\ref{fig3}b), which is the electrically-induced birefringence of non-centrosymmetric materials with $\chi^{(2)}$ nonlinearities\cite{karvounis2020electro, benea2021electro, weigand2021enhanced, benea2022gigahertz, zheng2024dynamic}. However, the refractive index modulation induced in this way is rather small, and conventional devices therefore generally require a large propagation length or a high quality factor response to achieve full phase or amplitude modulation depth. Using the resonant features of metasurfaces, one can substantially reduce the propagation length needed for full wavefront control. Modulation speeds up to several gigahertz have been demonstrated using the electro-optic Pockels effect with large bias voltages ($[-100,100]$~V)\cite{benea2022gigahertz} in a resonant configuration. An alternative electro-optic effect with lower power consumption is the quantum confined Stark effect\cite{kuo2005strong} with driving voltages potentially under 1~V and theoretical sub-picosecond switching speeds. This effect has been employed in optical metasurfaces with multiple quantum wells\cite{wu2019dynamic}, but only with partial wavefront phase control. 

Overall, the electro-optic effects described above are a promising route to implement optical neural networks, owing to their high cyclability and electrical tunability. The high switching speed of this mechanism gives it a competitive edge, particularly for the implementation of optoelectronic feedback loops, such as the structural nonlinearity mentioned above. A drawback of this technique is that it is limited to narrowband resonant responses, which precludes spectral multiplexing. 

\subsection*{Liquid-crystal metasurfaces}

\noindent
Liquid-crystal-infiltrated metasurfaces\cite{mikheeva2022space, gu2023reconfigurable} have also been used to perform bias-dependent beam-steering. Such metasurfaces exploit the sensitivity of liquid-crystal molecules to an externally applied electric potential and utilize their birefringence to modify the refractive index surrounding resonant meta-atoms. When an electric field is applied, the liquid-crystal molecules re-orient themselves so that they are aligned along the field lines\cite{khoo2022liquid} (for positive dielectric anisotropy molecules, see Fig.~\ref{fig3}c). The re-orientation commences at a threshold bias known as the Fréedericksz transition, beyond which the effective refractive index of the liquid-crystal medium varies as a function of the applied bias, for example, between the ordinary ($n_{\text{o}}$) and extraordinary ($n_{\text{e}}$) indices of refraction (see Fig.~\ref{fig3}c). All existing resonant meta-atom modulation schemes rely on the tuning of the surrounding refractive index to shift the meta-atom resonances. Although the index change induces a voltage-dependent phase modulation of the field traversing a pixel of the liquid-crystal metasurface, the resulting resonance shift leads to either a phase and/or amplitude modulation around a specific narrow bandwidth. Nonetheless, achieving broadband continuous phase tuning with unitary amplitude, either with the Pancharatnam-Berry or Huygens meta-atom design, is a challenge that has yet to be met. The use of liquid crystals in a nonresonant effective refractive index modulation mechanism could circumvent these issues.

Liquid crystals have been extensively used in electro-optic devices, such as displays or spatial light modulators, with long-term operation capabilities, low power consumption, and industrial maturity (see Table~\ref{table1}). Such technological maturity suggests that one can expect liquid-crystal metasurfaces to be among the first platforms used for future PNN devices. However, under certain input polarization conditions, the polarization dependence of liquid crystals could reduce the modulation efficiency. Additionally, the response time of liquid crystals in bulk devices can be long, which can be a limiting factor for optical computation. Recent works have been proposed to resolve this issue, mostly by relying on subwavelength liquid-crystal confinement and subwavelength voltage-actuated liquid-crystal unit cells\cite{lumo1, lumo2, kyrou2024fast}. Such geometries achieve full phase modulation while reducing the liquid-crystal actuation distance by about one order of magnitude, thus improving the switching speed by about two orders of magnitude. The subwavelength resolution and higher switching speeds would undoubtedly give such liquid-crystal devices a competitive edge in optical computing applications. Several questions arise, however, when implementing this unit cell geometry, including the issues associated with the liquid-crystal anchoring and the role/impact of a superficial liquid-crystal layer, which is often present at the top of the infiltrated nanostructures.

\subsection*{Phase-change materials}

\noindent
Another widely employed modulation mechanism in optical metasurfaces relies on phase-change materials (Fig.~\ref{fig3}d), mostly phase transition oxides and chalcogenides\cite{wuttig2017phase, ke2018vanadium, abdollahramezani2020tunable, cueff2020vo2}. This approach achieves a refractive index modulation by modifying the structural atomic arrangement of a material by heating or cooling either electrically or optically. The modulation mechanism based on phase-change materials is particularly interesting for reconfigurable optical metasurfaces because these materials provide a large refractive index modulation and a fast response\cite{ding2019dynamic, meng2020progress} (see Table~\ref{table1}). 

The different material platforms have different advantages. For example, phase transition oxide materials, which are typically used to tune the resonances of metasurfaces, operate at low temperatures but display high absorption, especially in the metallic state, whereas chalcogenides offer high cyclability, low power consumption, fast switching speed, and long retention of the state. However, the switching between the different atomic structures of chalcogenides requires heating to high temperatures around $ 600^{\circ}$C, which makes careful thermal design and packaging crucial for the co-integration of chalcogenide-based devices with digital hardware. Integrated platforms that embed phase change materials are suitable for at least basic linear optical computing\cite{dinsdale2021deep}, and their level of maturity indicates that PNN implementations with such materials are expected in the near future.

\subsection*{Other modulation schemes}

\noindent
Another modulation scheme involves the refractive index change induced by free-carrier density modulation\cite{shcherbakov2017ultrafast, howes2018dynamic}, which can either be controlled electrically or by optical pumping. This implementation remains limited because of absorption losses and faces additional integration challenges with digital architectures. Mechanical tuning\cite{chang2020metamaterials, chen2021electromechanically} and chemical reactions\cite{peng2019scalable, kaissner2021electrochemically} can also be employed for optical modulation. However, these approaches are ill-suited to individual element control, which is indispensable for arbitrary matrix-vector multiplication, and also suffer from extremely slow switching speeds  (see Table~\ref{table1} for a comparison of the performances and compatibility of the various modulation mechanisms).

\subsection*{Some implementations of reconfigurable metasurfaces in optical computation}

\noindent
Some efforts have been made to incorporate the benefits of reconfigurability into PNN hardware using only non-reconfigurable metasurfaces. In one such implementation, different static metasurface layers were mechanically translated to form different diffractive neural networks, thus enabling task switching\cite{he2024pluggable}. In another approach, a free-space optical data storage device was realized using the relative rotation of two holographic metasurfaces, allowing the storage of multiple images with only two compact layers\cite{fan2024holographic}. While effective for data storage, these strategies rely on macroscopic mechanical movement of static metasurfaces, limiting their speed and adaptability.\\

As an emerging field, the use of reconfigurable (or programmable) metasurface hardware for executing neural network tasks remains largely unexplored, with only a limited number of demonstrations reported in the literature. While these early implementations are still short of achieving the performance targets envisioned in this Perspective, they serve as important proofs of concept and offer valuable insights into key challenges the community faces to advance this technology.\\

The earliest reported use of reconfigurable metasurfaces for neural network computation dates back to 2021, when a waveguide mode converter incorporating a phase-change GeSbTe metasurface was employed to implement a convolutional neural network for edge detection and image recognition\cite{wu2021programmable}. By toggling the GeSbTe material between amorphous and crystalline phases, the input TE\textsubscript{0} mode was partially converted to the TE\textsubscript{1} mode in a controllable way, achieving up to 64 discrete levels corresponding to different crystallization states. The modal composition, which was detected using two photodiodes, effectively resulted in a multiplication operation over the incident light. While this device employed a metasurface-like structure for mode conversion, it more closely aligned with the PIC paradigm, where light propagates in-plane. Notably, the relatively large footprint of each functional unit (exceeding 10~$\mu\text{m}^{2}$ per operation) highlights a key limitation in terms of scalability.\\

Other implementations of reconfigurable metasurfaces have focused on related but distinct computational tasks, particularly within the field of optical image processing. The tunability of metasurfaces has been exploited to dynamically switch between different processing tasks, including differentiation for edge detection, angular filtering, and bright field imaging. Various reconfiguration strategies have been used, including the use of phase-change materials\cite{cotrufo2024reconfigurable} and mechanical stretching of the substrate\cite{zhang2021reconfigurable}. In both cases, the metasurfaces are designed to compute the Laplacian of the input image by engineering their angular transmission dependencies --- an approach often referred to as Green’s function engineering. These implementations rely on a global tuning mechanism to activate or deactivate the processing function, hence avoiding the complexity of controlling individual metasurface elements. However, the absence of subwavelength-level tunability imposes fundamental limitations on the flexibility of task-specific functionalities. Achieving fast independent pixel-level control is the next milestone for advancing metasurfaces and their application in optical computing.

\section*{Potential of programmable metasurfaces for photonic AI}
\noindent
The key advantage of alternative photonic computing hardware, especially free-space programmable metasurfaces, is their potential to overcome the challenges faced by digital systems, particularly in terms of processing speed and energy efficiency in large-scale models\cite{colburn2019optical, huang2024photonic} (see also Box~\ref{box 3}). In addition, having a device in which optical and electrical signals perform the computation and the modulation, respectively, provides an alternative route to computing devices beyond the von Neumann architecture. 

This architecture is susceptible to the so-called von Neumann bottleneck, which arises because both program data and instructions are transferred sequentially from the same memory unit over the same bus. This severely limits the throughput between the central processing unit (CPU) and the memory, particularly for compute-intensive applications requiring frequent memory access. In addition, as the scale and complexity of computations increase, so does the energy consumption associated with moving data between the CPU and memory, leading to inefficient energy usage. 

Despite its potential for energy-efficient operation, general-purpose optical computing hardware remains elusive because there are no high-performance optical analogues of electronic transistors\cite{miller2010optical}. Similarly to quantum computers, therefore, it is generally accepted that optical computers are best suited to special-purpose applications in which computational accuracy is not a critical requirement\cite{mcmahon2023physics, li2024exploring}. This limitation also applies to the metasurface approach discussed here. 

Nevertheless, the reconfigurable metasurface approach shows promise in delivering enhanced performance compared to other photonic platforms, as illustrated in Box~\ref{box 3}. Beyond these improvements, metasurfaces have well-established multiplexing capabilities, including vectorial wavefront shaping, orbital angular momentum, and simultaneous multi-wavefronts, as well as wavelength multiplexing, which can be advantageous for neural network implementation (Fig.~\ref{fig4}a). Programmable metasurfaces afford an additional multiplexing dimension since each pixel can be activated either using a DC or an AC voltage. The use of an AC voltage induces signal harmonics at the output of the metasurface\cite{zhang2018space} 
, a peculiarity that can be exploited to simultaneously encode multiple outputs at varying harmonics of the modulation frequency for the same input (see Fig.~\ref{fig4}b), thus endowing the neural network with multitasking capabilities.

We note here the influence of subwavelength meta-atoms on optical computation. The use of sub-wavelength pixels enhances energy efficiency by eliminating losses associated with undesired diffraction orders. Additionally, most reconfigurable liquid-crystal metasurface implementations resolve the issue of strong fringing effects arising at the interface between adjacent pixels in liquid-crsytal on silicon SLM devices. A well-known result from Fourier optics is that the spread in spatial frequency components from an aperture increases with decreasing aperture size. Subwavelength pixels thus enable a higher angular spread, yielding better pixel-to-pixel wide-angle connectivity between two consecutive layers, and could reduce the thickness of the overall device\cite{miller2023optics, li2024spatial}. For example, the minimum thickness needed for an optical device to perform a given optical computation is\cite{miller2023optics} $C\lambda/2n(1-\cos \theta )$, where $\lambda$ is the wavelength of light, $C$ is the number of channels that must pass from input to output, $n$ is the largest refractive index in the device, and $\theta$ is the angle giving the device's numerical aperture. Consequently, this lower bound on device thickness is reduced in higher numerical aperture devices. 

The higher resolution also enables PNNs to mimic complex optical systems that match the resolution of high-end detection modules and CCD cameras, in which the output signal is encoded down to sub-micron resolution. Note that in PNNs, the resolution of the input/output layers can differ from that of the hidden diffractive layers, particularly in convolutional neural networks. The subwavelength pixel size also enables the development of ultra-compact large-parameter neural network models. As a final benefit, recent reconfigurable metasurface implementations have also demonstrated significantly higher modulation bandwidth (up to several GHz, see Table~\ref{table1}) compared to conventional free-space wavefront modulation devices.

Reconfigurable metasurfaces with subwavelength resolution enable the following key advantages for PNN applications. 
\begin{itemize}
    \item Multitasking optimization with a single device: The reconfigurable metasurface neural network can be optimized for multiple inference tasks, as the tunable pixels of the neural network can modify the connections based on the task to be implemented. Multitasking is an essential component of artificial intelligence, which is still largely lacking in photonic AI hardware. 
    \item Structural nonlinearities: Nonlinearities can be implemented by repeatedly feeding input data --- or any user-defined transformation of them --- into each layer. Note that, to date, there is no standard method to include nonlinearities in PNNs. Structural nonlinearities are thus a promising route to standardizing nonlinear operation in photonic AI. 
    \item Real-time learning: The network supports real-time or in situ learning, allowing training and inference to occur at the same time for faster adaptation. This is critical for enabling advanced training schemes such as physics-aware training, which is impossible to implement in a non-programmable device.
    \item Reduced device thickness: The subwavelength pixel size increases the angular spread of light, such that the minimum thickness needed to perform a given optical computation is reduced\cite{miller2023optics}. This would translate to more compact devices that can be better integrated with other hardware.
    \item Robust to fabrication errors: The fact that reconfigurable metasurface PNNs consist of meta-atoms whose electromagnetic properties can be continuously tuned and optimized for a specific task makes such devices highly robust to fabrication imperfections. Note that fabrication errors can negatively impact performance in typical optical computing structures. 
\end{itemize}

\subsection*{Promising metasurface architectures}

\begin{figure*}[ht!]
\centering
\includegraphics[width=\linewidth]{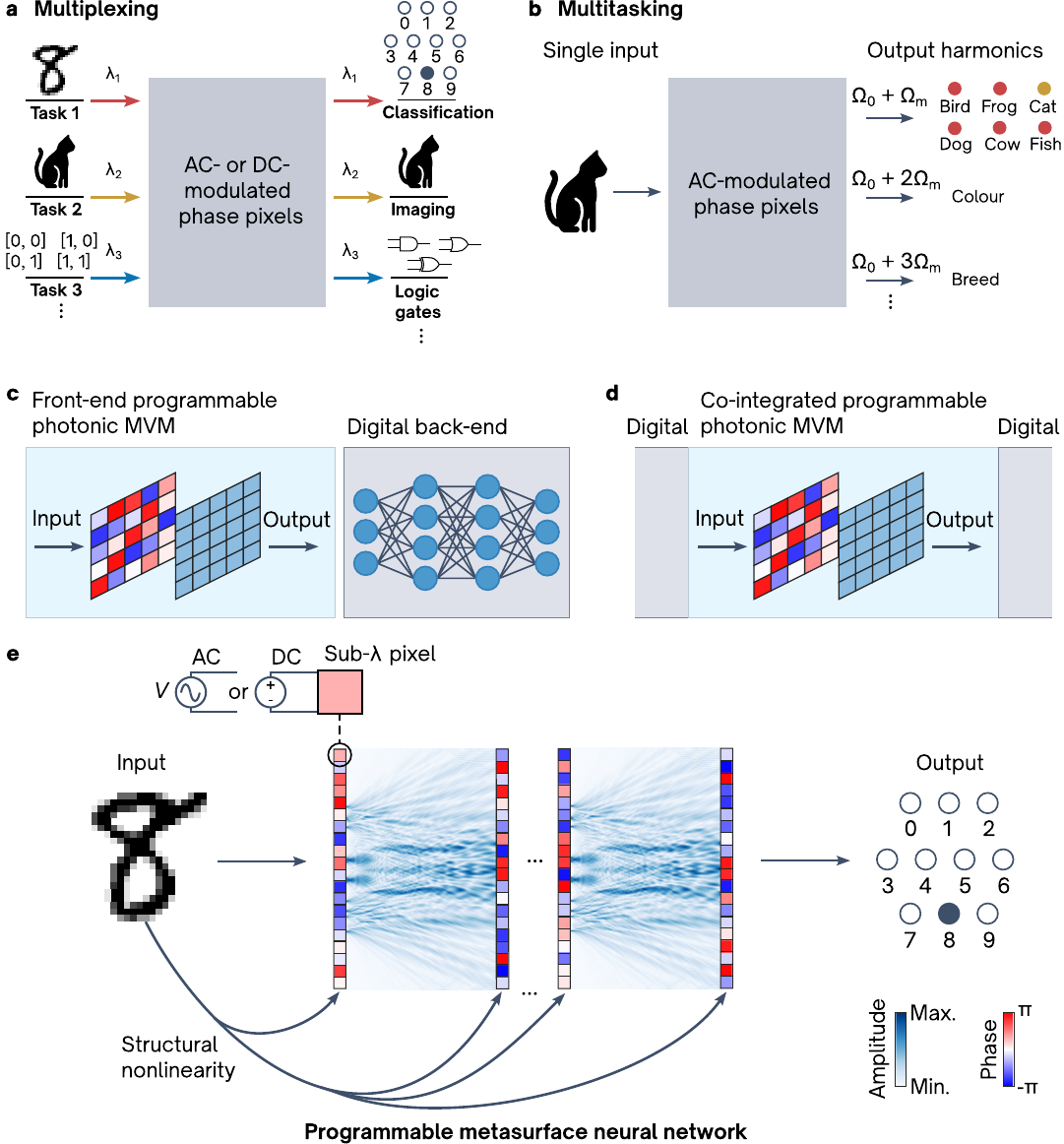}
\caption{\textbf{Possible neural network architectures with programmable metasurfaces.} Potential neural networks implemented with programmable metasurface matrix-vector multipliers (MVMs), which rely on sub-wavelength pixels that can be actuated either with a DC or an AC electrical bias, enable \textbf{a}, wavelength multiplexing and \textbf{b}, multitasking. The multitasking capability arises from the AC modulation that induces signal harmonics, each of which is used to encode a different output. \textbf{c}, A schematic illustration of a programmable metasurface MVM front-end-integrated and \textbf{d}, co-integrated as an optical AI accelerator to a digital neural network. \textbf{e}, Schematic of a stacked programmable metasurface MVM neural network subjected to structural nonlinearity by superposing the input information at each layer.}  
\label{fig4}
\end{figure*}

\noindent
As computing metasurface devices need to integrate and be compatible with existing flexible and robust digital technology, we expect them to be implemented in hybrid optical/digital hardware. For example, a programmable optical metasurface can be integrated at the front-end of a digital computer to optically perform an arbitrary kernel on the input data (see Fig.~\ref{fig4}c). This approach has the advantage of pre-processing the data in the optical domain, effectively bypassing one digital-to-analogue conversion as the input data are composed of optical signals. Such front-end integration has also been shown to reduce the dimensionality of data, which unloads part of the computation from the digital back-end\cite{xia2023deep, zheng2024multichannel, swartz2024broadband}. The programmable metasurface layer can also be integrated with a digital computer as an optical accelerator (Fig.~\ref{fig4}d), converting digital data to optical signals, performing some optical computation, and then converting the result back to the digital domain. 

We have only mentioned the most common possible architectures here, but a complete 'catalogue' of architectures can be envisioned, with the programmable diffractive layer (or layers) intervening at a different section of the hybrid digital-optical neural network device. One can imagine the diffractive layer(s) integrated with a convolutional neural network, implementing only the convolution layers at the beginning of the pipeline or only the fully connected layers terminating it. Alternatively, these diffractive layers could be used in a recurrent neural network  architecture\cite{zhou2021large}, where the depth of the network can be increased. Other architectures have been discussed in the literature, such as residual and graph neural networks and variational auto-encoders\cite{brunner2025roadmap}, which require assessing the best-suited use of diffractive layers with or without nonlinearities.  

Ultimately, the progress in metasurface manufacturing and integrability will enable the stacking of several reconfigurable layers (see Fig.~\ref{fig4}e), which can be updated on demand to perform various computational tasks. This scheme has the benefit of supporting nonlinear activation through structural nonlinearities, as previously mentioned. Aside from the manufacturing difficulties of such a standalone device (as well as co-integrated devices), other challenges may arise, such as the requirement of developing suitable workflows for training and deploying models specifically designed for PNNs. This, in turn, brings additional complexities associated with the adoption of such reconfigurable technology by software engineers, switching costs, and the modification of the programming and training paradigm developed for digital computing. To mitigate the acceptability risk of introducing this new technology, simplistic photonic neural networks would initially work with current digital hardware, much like the CUDA library was developed to use GPUs in regular computing. Alongside these difficulties emerge numerous opportunities to establish new fields of research and development, much like the ecosystem that formed around the paradigm of quantum computing in the past.

\section*{Open challenges}
\noindent
Before optical computation based on reconfigurable metamaterials can be deployed in commercial AI applications, several challenges must be addressed. The most critical obstacle is scaling up fabrication while maintaining precise control over the device's many
parameters. This issue can be resolved by exploiting the fabrication capabilities of complementary metal-oxide semiconductor (CMOS) foundries. Despite the maturity of CMOS processes, only a few have been specifically adapted for the fabrication of programmable metasurfaces. Nonetheless, some of the developed CMOS fabrication steps for electronic chips (logic, memory, and processing) share the same technological requirements as those of programmable metasurfaces, such as the need for high aspect ratio etching in memory chips\cite{ishikawa2018progress} or dense interconnectivity with via-holes in processing chips. Early-stage adaptations have already started; for example, CMOS fabrication processes have been implemented with phase-change materials\cite{arnaud2018truly} and non-reconfigurable metasurfaces\cite{lepers2023metasurface, vaillant2024metasurface}. The most crucial bottleneck limiting the full adaptation of CMOS processes for reconfigurable metasurfaces is the existence of a commercially viable application that would justify substantial development costs; optical computing hardware could be such an application. 

Another obstacle is related to the individual control of the meta-atoms, which is necessary for the full control of the associated wavefront. The current approach for electrical biasing relies on the in-plane redistribution of connectors, which is limited to a few tens of individually controlled elements\cite{li2019phase, park2021all}. A more scalable scheme involves connecting each meta-atom from the substrate using via-holes, which consist of conductive vertical connections used to electrically bias individual pixels. Further complexity with the via-hole approach arises in the case of volatile pixels that lose the desired optical response once the external stimulus is removed, thus requiring continuous bias. This is particularly the case for electro-optic effects and liquid crystals, where active matrix addressing schemes are employed in which supplementary circuitry (transistors and capacitors) is used to maintain the bias at each pixel. The requirements of via-hole biasing and additional circuitry suggest that the device's photonic and electronic modules have to be considered in concert at the design stage. \\

Achieving such co-integration of nanocomponents requires advanced foundry manufacturing and close coordination during the design phase. This, however, is not a critical limitation as such co-integration, specifically in the case of the active matrix addressing scheme, has already been implemented commercially, for instance in liquid crystal on silicon SLMs, but with rather larger pixel size (or pixel pitch). Current CMOS backends can already support pixel sizes down to $\sim 1\,\mu$m in advanced nodes, and progress in nanofabrication, driven by stacked architectures and improved deep trench isolation, is expected in the coming $1-3$ years. We anticipate that this trend would ultimately reduce the pixel addressing size down to the subwavelength scale ($0.5-0.7 \,\mu$m), which is in the regime needed for this technology. 

Current non-programmable optical computation technology has already been integrated into photonic devices with digital hardware, demonstrating certain advantages over digital computation, including intrinsic massive parallelism, high processing speed, drastically lower energy consumption\cite{colburn2019optical, huang2024photonic}, ultra-high bandwidth, and low latency. However, novel optical computing architectures, such as the ones discussed in the previous subsection, which can take full advantage of optical reconfigurability, still need to be developed and assessed before the mass adoption of this technology. 

In analogy with digital field-programmable gate arrays, we envision a game-changing technology, which we refer to as a field-programmable metasurface array (FPMA), a photonic device capable of changing its computing function on the fly, for example, by switching between arbitrary matrix-vector multiplications or even modifying its network architecture without the costly need of re-fabricating the hardware. Operations or transformations such as Fourier transforms, convolutions, and so forth can be dynamically changed, enabling adaptive computation, that is, switching between wave-based computation modalities with a single architecture.  FPMAs would achieve multi-terabits/sec throughput by using true parallel processing, which is of interest for image processing and graph computation, and multiplexing capabilities in various dimensions, such as in wavelength, phase, amplitude, polarization or angular momentum.

Reconfigurable metasurfaces appear as the ideal optical devices for advanced AI and neuromorphic computing, performing matrix-vector multiplication using built-in interference and diffractive optical properties, provided that CMOS supply chains are adapted to this technology and dedicated architectures and software are developed. Additionally, metasurface technology is naturally compatible with quantum optical information processing, which may be exploited to bridge classical-optical and quantum-optical computing. Programmable metasurfaces can also enable the "post-von Neumann computing era", in which the same structure uses different buses to send and compute information directly via light signals, executing massive matrix operations as light propagates through the devices. Electronic modulation at each of the programmable metasurface's pixels is the instruction set, and light interacting with the pixels is the data path.

The future of smart optics is bright, particularly with the development of field-programmable metasurface arrays, which provide a wave-based computing platform that yields adaptive and real-time in-place computation, eventually challenging the speed, energy efficiency, and flexibility of traditional digital architectures. 


\section*{Acknowledgments}

\noindent
E.M. acknowledges support from the Marie-Curie Common laboratory program H2020  MSCA-RISE-2020, Project CHARTIST, under Project Number 101007896.

\section*{Author contributions}

\noindent
L.A.H. compiled the figures. E.M. compiled the table giving the performance metrics of the discussed optical modulation schemes. P.G., L.A.H., and E.M. conceptualized the different light modulation schemes discussed in the manuscript. P.W. wrote the numerical code for the implementation of the diffractive XOR logic gate. L.A.H., E.M., and P.G. wrote the manuscript with contributions from P.W., P.d.H., and T.W.

\section*{Competing interests}

\noindent
The authors declare no competing interests.

\section*{Text availability and additional information}

\noindent
The main text and supplementary information of this article are available here: \url{https://doi.org/10.1038/s42254-025-00831-7}

\vspace{0.25cm}

\noindent
A view-only PDF of the published article is available here: \url{https://rdcu.be/elEmz}


\bibliography{ref}

\newpage


 \begin{table*}[h]
\centering
\resizebox{\linewidth}{!}{
\begin{tabular}{ccccccc}
    \hline
    \hline
    & \textbf{Electro-optic} \quad & \textbf{Liquid-crystals} \quad & \textbf{Phase-change} \quad & \textbf{Free-carrier} \quad  & \textbf{Mechanical} \quad & \textbf{Chemical} \quad \\
    & \textbf{modulation} \quad & \quad & \textbf{materials} \quad & \textbf{density modulation} \quad & \textbf{tuning} \quad & \textbf{tuning} \quad\\
    \hline
    \textbf{Bias type} & Voltage
 & Voltage
 & Electrical or optical  & Electrical or optical  & Mechanical  
 & Voltage \\
 & & & heating & pumping & stress &  \\
    \hline
    & & & & & & \\
    \textbf{Modulation} & up to 5~GHz\cite{benea2021electro}
& $>500$~Hz* 
~\cite{kowerdziej2019ultrafast} 
 & $10\, \text{kHz}$\cite{wang2021electrical}-$10\, \text{MHz}$\cite{zhang2019designing} & few tens of GHz\cite{lee2014nanoscale}
 & $1-200$~kHz\cite{chen2021electromechanically} 
 & $<100$~Hz\cite{huang2019voltage, kaissner2021electrochemically} \\
    \textbf{bandwidth} & & & &  &  & \\
    \hline
    \textbf{Optical efficiency} & Up to 100\%\cite{benea2021electro} &  $<65$\%*~\cite{kowerdziej2019ultrafast, sun2019efficient} & $25-50$\%*\cite{gholipour2013all, wang2021electrical} & $0.5-35$\%*\cite{lee2014nanoscale, shcherbakov2017ultrafast}
 & Up to 90\%\cite{chen2021electromechanically}  & $\sim 100$\%\cite{kaissner2021electrochemically} \\
    \hline
        \textbf{Power consumption}$^{\dag}$ & High voltage  & Low voltage & $<25$~mW/cm (optical)\cite{gholipour2013all} & $\sim 0.2$~W (optical)\cite{shcherbakov2017ultrafast}  & High voltage & $\sim 5$~V\cite{huang2019voltage}  \\
     \textbf{of modulation} & $-100-100$~V\cite{benea2021electro, zheng2024dynamic}  & $1-5$~V & $>6$~kW* (electrical)\cite{wang2021electrical}  & $\sim 15$~mW (electrical)\cite{lee2014nanoscale} & 10s of V\cite{chen2021electromechanically} & \\
    \hline \\
    \textbf{Cyclability} & $> 10^{12}$ & $10^{9}-10^{12}$ & $\sim 10^{15}$ & $> 10^{12}$ & $\sim 10^{11}$ & $\sim 100$ \\
     & cycles & cycles & cycles\cite{yang2021switchable} & cycles &  cycles\cite{dahl2020effect} & cycles\cite{huang2019voltage}  \\
    \hline
    \textbf{Temperature} &Very & Up to $140^{\circ}$C\cite{gauza2012high} & Up to $140^{\circ}$C\cite{wang2021electrical} & Very & Room & \\
    \textbf{stability} & robust\cite{zheng2024dynamic} & & &  robust & temperature & Up to $200^{\circ}$C\cite{chandrakanthi2000thermal} \\
    \hline
    \hline
\end{tabular} }
\caption{\textbf{A comparison of various modulation techniques based on key performance metrics.} The reader is referred to Refs.~\cite{mikheeva2022space, gu2023reconfigurable} for a more thorough review of reconfigurable optical metasurfaces. * This figure is based on current implementations, but because of extensive research on this subject, this figure is expected to improve in the near future. $\dag$ Power consumption values are given in watts when the value is disclosed or readily computable from the cited studies. Often, however, only the biasing voltage is disclosed, and the equivalent electrical circuit that would enable the computation of the actual power consumption is not straightforward. In these cases, we give a bias voltage as an indicative value. In general, lower voltages correspond to lower power consumption.}
\label{table1}
\end{table*}

\newpage


\newpage


\begin{figure*}

\begin{mdframed}[backgroundcolor=gray!10, linecolor=black, frametitle={\boxtitle{The massive energy consumption of modern digital neural networks}\label{box 1}}, frametitlebackgroundcolor=gray!30, frametitlealignment=\center]

   \vspace{0.25cm}

\noindent
\begin{minipage}[t]{0.48\textwidth}
\justifying
 Energy consumption associated with memory access, data transfer, and performing operations in digital neural networks is staggering. As a result of the excessively large number of floating-point operations per second\cite{openai} in both the training and inference phases of deep digital neural networks, energy consumption is likely to continue growing as the complexity of digital models increases. During the training phase, these floating-point operations consist of passing a portion of a rather large dataset through the network in a feedforward pass, followed by a back-propagation of the errors\cite{rumelhart1986learning}. The inference phase relies on a single forward pass and, in principle, consumes less energy, but it could still become substantial in some cases.

  To illustrate this issue, we estimate the energy consumed during the training phase of state-of-the-art large-language models, namely Meta's Llama 3 and OpenAI's ChatGPT-4. According to Meta, Llama 3 was trained on a 24,576-NVIDIA H100 graphics processing unit (GPU) data centre-scale cluster\cite{meta}, with an average power consumption of 700~W per GPU. Although undisclosed by Meta, a good estimate of the duration of the training process is around 90 days (2,160 hours), leading

\end{minipage}%
\hspace{0.04\textwidth}%
\begin{minipage}[t]{0.48\textwidth}
\justifying
\noindent
  to a total energy consumption of 
 \begin{align}
      \mathrm{Energy \, consumption} &= 0.7\, \mathrm{kW} \times 24,576 \times 2,160 \, \mathrm{hours}\nonumber \\
      &= 37.16 \times 10^{6} \, \mathrm{kilowatt-hours} .\nonumber 
\end{align}
 \noindent
   The energy consumed during the training phase of Llama 3 is thus a whopping 37.16~GWh. A similar estimate can be obtained for the training phase of ChatGPT-4. For reference, the average annual energy consumption of an American household is around 10 MWh\cite{UShouse}, and that of a desktop PC (used 8 hours daily) is between $175-730$~kWh\cite{deskpc}. It should be emphasized here that, even after training, the energy required to operate these LLMs remains considerable, with an estimated energy consumption of 5~J per token generation in Llama 65B (a previous-generation LLM from Meta), for example\cite{samsi2023words}. Assuming an average user generates tens of thousands of tokens daily, delivering service to a million users would require daily electric energy on the order of tens of megawatt-hours. We also note that the figures computed above are usually multiplied by a power usage effectiveness factor, which is typically between 1.2 and 2 (see ref.~\cite{google2023efficiency}), accounting for that part of a data centre's energy usage that is indirectly related to computing (such as cooling and overhead).
\end{minipage}

\vspace{0.25cm}

\end{mdframed}
 \end{figure*}


\begin{figure*}

\begin{mdframed}[backgroundcolor=gray!10, linecolor=black, frametitle={\boxtitle{Multi-layer diffractive elements as accelerators of deep neural networks}\label{box 2}}, frametitlebackgroundcolor=gray!30, frametitlealignment=\center]

   \vspace{0.25cm}
   
\noindent
\begin{minipage}[t]{0.48\textwidth}
\justifying
Matrix-vector multiplication accounts for the majority of operations required by neural network models and, thus, for the majority of memory and energy usage. Multi-layer diffractive elements have been proposed as a scalable solution for unloading matrix-vector multiplication costs in digital neural networks, that is, as optical AI accelerators. In this box, we draw a parallel between deep digital neural networks and multilayer diffractive elements, highlighting how the latter are used as optical matrix-vector multipliers.

The basic scheme of a layered deep digital neural network (panel \textbf{a} below) consists of an input layer at the top, any number of intermediate (hidden) layers, and a layer of output units at the bottom. Any number of inter-layer connections can be formed in a descending fashion (from top to bottom), and, in the simplest case, intralayer connections are forbidden. An input vector is presented to the network by setting the states of the input units. The network's layers have their states set sequentially, starting from the top and working downwards until the states of the output units are determined. The total input vector to layer $l+1$, $\mathbf{X}_{l+1}$, consists of the matrix multiplication between the output vector of layer $l$ ($\mathbf{X}_{l}$) and the weight matrix $\mathbf{W}_{l}$, to which a bias $\mathbf{B}_{l}$ is added. Finally, a nonlinear activation function $\mathbf{F}_{l}$ is applied to the sum of all input values to each neuron (typically, this is a rectified linear-unit function). Formally, this operation is expressed as $\mathbf{X}_{l+1} = \mathbf{F}_{l}(  \mathbf{W}_{l}\mathbf{X}_{l} + \mathbf{B}_{l})$ (see panel \textbf{c} below). The learning process is realized by optimizing each of the coefficients of all the weight matrices applied between the layers during the training phase only so that the neural network can closely yield the desired outputs based on a well-approximated function.
\end{minipage}%
\hspace{0.04\textwidth}%
\begin{minipage}[t]{0.48\textwidth}
\justifying
\noindent

A diffractive matrix-vector multiplier (panel \textbf{b} below), on the other hand, is constructed from multiple pixelated photonic layers, in which each pixel on a given layer has a complex-valued transmission (or reflection) coefficient, which is optimized so that each layer performs a specific matrix-vector multiplication. In the absence of multiple scattering between layers and mutual coupling within layers and following the Huygens-Fresnel principle, each pixel on a given layer acts as a secondary source of a wave, whose amplitude and phase are determined by the product of the input wave and the complex-valued transmission (or reflection) coefficient at that point. Consequently, the weights on the connections between the various neurons are modulated in amplitude and phase owing to the interference of all secondary waves propagating between the layers, and the complex transmission (or reflection) coefficient of each neuron. The transmission coefficient of each neuron acts as an element-wise scaling on its connections with the neurons of the preceding layer. The mathematical operation performed between the network's layers is $\mathbf{Y}_{l+1} = \mathbf{W}_{l+1} (\mathbf{Y}_{l} \, \mathbf{o} \, \mathbf{B}_{l})$, in which $\mathbf{Y}_{l} = \mathbf{X}_{l} e^{i\bm{\phi}}$ is the input to layer $l$, $\mathbf{B}_{l}$ is the bias term corresponding to the complex transmission coefficient of layer $l$, and the $\mathbf{o}$ symbol denotes element-wise matrix multiplication (see panels \textbf{b} and \textbf{d} below). Note that in this case, a non-linear activation function (other than the read-out nonlinearity of the detector at the output layer) is absent, which suggests that exploring nonlinearities in such structures could be an intriguing research direction. 

\end{minipage}

\begin{center}
\includegraphics[width=0.75\linewidth]{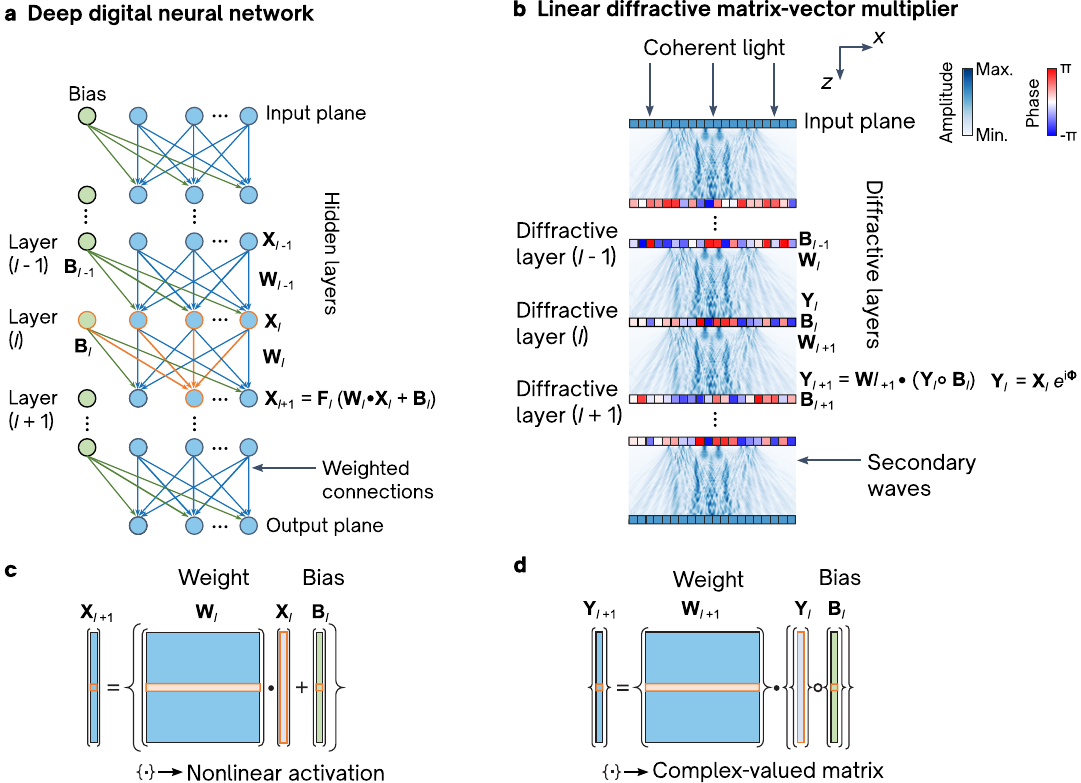}
        \captionof*{figure}{\textbf{Deep digital neural networks and linear diffractive matrix-vector multipliers.} \textbf{a}, Schematic illustration of a deep digital neural network and \textbf{b}, a diffractive matrix-vector multiplier, along with the mathematical operations occurring between the cascaded layers, panels \textbf{c} and \textbf{d} respectively. The "\textbf{o}" symbol denotes a Hadamard product operation (element-wise multiplication). The electric field amplitude shown between the layers in panel \textbf{b} is the diffracted field of an arbitrary image (here we used the Colorado School of Mines' affiliation logo) whose propagation in the longitudinal direction was numerically calculated using the angular spectrum method, implemented via a Python package\cite{testurl}. [Panels \textbf{a}, \textbf{c}, and \textbf{d} adapted from Ref.~\cite{lin2018all}.]}
    \end{center}

\vspace{0.25cm}
\end{mdframed}

\end{figure*}


\begin{figure*}
    
\begin{mdframed}[backgroundcolor=gray!10, linecolor=black, frametitle={\boxtitle{Performance metrics of electronic and optical computation}\label{box 3}}, frametitlebackgroundcolor=gray!30, frametitlealignment=\center]

   \vspace{0.25cm}

\noindent
\begin{minipage}[t]{0.48\textwidth}
\justifying
This box provides a comparison of the performance metrics of 4 hardware platforms for neural network computing, namely, graphics-processing units (GPUs) with complementary metal–oxide–semiconductor (CMOS) electronics, photonic-integrated circuits, liquid-crystal on silicon spatial light modulators (LCoS SLMs), and future programmable optical metasurfaces. 

The key metrics we have chosen here are component density (CD) in cm$^{-2}$, compute speed in tera ($10^{12}$) operations per second (TOP/s), and energy per operation (pJ/OP). These values are summarized in the table below. The listed component densities for the LCoS SLM and programmable metasurface were calculated by considering a pixel size of $10\times 10$~$\mu$m$^{2}$ and $300 \times 300$~nm$^{2}$, respectively. The compute speed of the photonic hardware is evaluated from the number of operations conducted by an $N\times N$-component network performing an $N \times N $ by $N \times N $ matrix multiplication and an $N \times N $ by $N \times N $ element-wise multiplication (see Box~\ref{box 2}).
\end{minipage}%
\hspace{0.04\textwidth}%
\begin{minipage}[t]{0.48\textwidth}
\justifying
 \noindent
  The matrix multiplication consists of $N^{3}$ multiplications and $N^{2}(N-1)$ summations, and the element-wise multiplication consists of $N^{2}$ multiplications so that the total number of operations is on the order of $N^{3}$. Thus, we have

\begin{equation}
\mathrm{Compute\, speed\, (TOP/s}) = \frac{\big(\sqrt{\mathrm{CD} \times 1 \, \mathrm{cm}^{2}}\big)^{3}}{10^{12}\times 
\tau }, \label{CD}
\end{equation}

\noindent
where $\tau$ is a time constant in seconds, giving the response time of the photonic neural network. We assume a response time of 1~$\mu$s for the photonic-integrated circuit network and 1~ms for the LCoS SLM and the metasurface. The energy per operation is computed from the ratio of the overall power of operation in watts to the compute speed in operations per second (OP/s). The operation power for a CMOS GPU was assumed to be\cite{H100} 700~W, and that of the metasurface was estimated to be 200~W. The energy per operation for LCoS SLMs and photonic-integrated circuits were taken from Ref.~\cite{zhou2021large, bandyopadhyay2024single}, respectively.
\end{minipage}

\vspace{0.25cm}

\begin{center}
\resizebox{0.9\textwidth}{!}{
    \begin{tabular}{ccccc}
    \hline
    \hline
         & CMOS & Photonic-integrated  & LCoS  & Programmable \\
        & electronics & circuits & SLM &  optical metasurfaces \\
        \hline
        Component density (cm$^{-2}$) & $10^{10}$ (Ref.\cite{H100}) \quad & $10^{3} - 10^{6}$ (Refs.\cite{bandyopadhyay2024single, bogaerts2020programmable}) \quad & $10^{6}$ &  $10^{9}$\\
   \hline \\
    Compute speed (TOP/s)& $2000$ (Ref.\cite{H100}) & $31.6-10^{3}$ & 1 & $3.16\times10^{4}$ \\
    \hline
     Energy per operation (pJ/OP)& 0.35 & $\sim 10$ (Ref.~\cite{bandyopadhyay2024single}) & 1.4 (Ref.~\cite{zhou2021large}) & 0.0063 \\
    \hline 
       \hline
       \end{tabular}
}
  \captionof*{table}{\textbf{A comparison of the performance metrics of electronic and photonic computing architectures.} }\label{table CD}
\end{center}
\vspace{0.25cm}
\end{mdframed}

\end{figure*}


\begin{figure*}

\begin{mdframed}[backgroundcolor=gray!10, linecolor=black, frametitle={\boxtitle{The role of non-linearities: The XOR logic gate example}\label{new box}}, frametitlebackgroundcolor=gray!30, frametitlealignment=\center]

   \vspace{0.25cm}

\noindent
\begin{minipage}[t]{0.48\textwidth}
\justifying
In this box, we give an example of a task in which a fully linear matrix-vector multiplier fails to produce the expected output, thus underpinning the role of nonlinearities in complex mathematical operations.

The 2-bit exclusive OR (XOR) logic gate operation is a standard example of a nonlinear binary logic operation that requires nonlinear activation to be implemented via a matrix-vector multiplier (MVM). This is because the 2-bit XOR gate takes two inputs and produces a single output, either True (1) or False (0), in which the True output arises if and only if the two input bits are different. As such, it cannot be represented as a linear combination of the inputs, making it a nonlinear binary logic operation. To implement the XOR operation with an MVM, at least one nonlinear activation function is needed.

A multilayer diffractive MVM can be trained to perform the XOR operation via error back-propagation learning.  As in Box~\ref{box 2} (panel \textbf{b}), the layers of the MVM consist of an array of phase pixels with values $\in [-\pi, \pi]$ that can be optimized to perform the XOR logic.
\end{minipage}%
\hspace{0.04\textwidth}%
\begin{minipage}[t]{0.48\textwidth}
\justifying
\noindent
 As illustrated in the panels below, the inputs are encoded into the phase images of the MVM, where an input of 0 is represented by uniform maximal phase (a phase value of $\pi/2$ across the entire image) and an input of 1 is represented by a focused Gaussian spot with minimal phase at its centre (a phase value of $-\pi/2$). A True output is represented by a Gaussian phase modulation spot on the left-hand side of the output phase image, while a False output results in a Gaussian phase modulation spot on the right-hand side of the output image. The input and output images are then plotted on a colour scale ranging from $-\pi$ to $\pi$ to avoid phase wrapping.

To compare the performance of linear and nonlinear activated networks, two multilayer diffractive MVMs were digitally set up and trained to perform the XOR operation, one with structural nonlinearities\cite{yildirim2023nonlinear} induced through the repeated re-injection of the input data at each layer and the other representing a fully linear MVM. As shown in the panels below, the diffractive MVM with structural nonlinearities (panel \textbf{a}) correctly produces the expected XOR output (compare the middle and bottom rows), whilst the fully linear MVM (panel \textbf{b}) produces erroneous outputs. Note that the simple XOR demonstration illustrates the limitation of linear systems but is not a typical example of a neural network inference task. 
\end{minipage}

\begin{center}
\includegraphics[width=\linewidth]{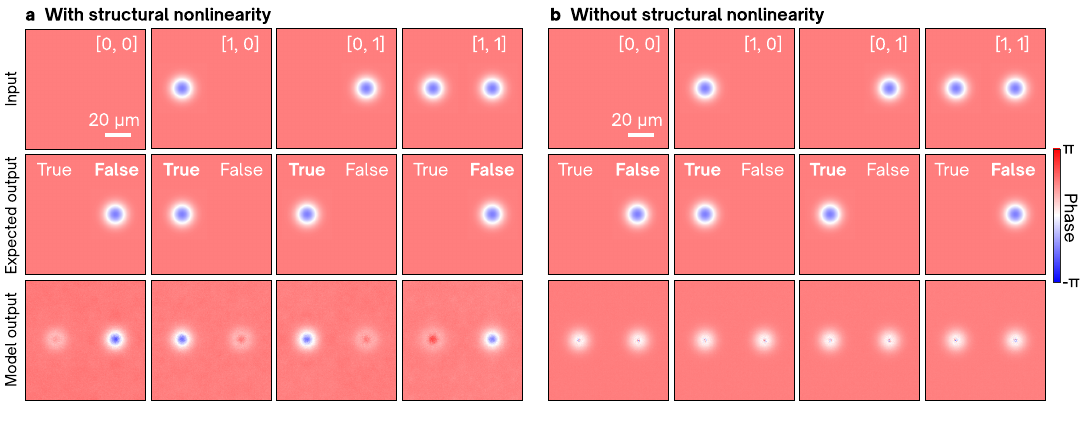}
        \captionof*{figure}{\textbf{An XOR logic gate implemented with a diffractive MVM.} A diffractive multi-layer MVM is digitally trained using error back-propagation to perform the XOR logic gate operation (see supplementary information, section~III for more details). The images correspond to the input and output phase images of the multi-layer phase-pixel MVM with the model output after training given in the bottom row. The results shown here correspond to an MVM consisting of four $294\times 294$ pixel equidistant layers (the separation distance is 100~$\mu$m) of size $100\times 100$~$\mu\text{m}^{2}$. The pixel size is $\lambda/2$, where $\lambda = 680$~nm is the wavelength of operation. Panel \textbf{a} gives the results of an MVM trained with structural non-linearities, while the results of panel \textbf{b} correspond to a fully linear MVM.}
    \end{center}

\vspace{0.25cm}
\end{mdframed}

\end{figure*}

\end{document}